\documentclass[12pt]{JHEP3}
\usepackage{amsmath,amssymb,amsfonts}
\usepackage[dvips]{graphicx}
\usepackage{epsfig}

\makeatletter \@addtoreset{equation}{section} \makeatother

\preprint{}

\title{\Large \bf BPS Electromagnetic Waves on Giant Gravitons}

\author{Seok Kim\\
\\
School of Physics, Korea Institute for Advanced Study, Seoul 130-012 KOREA\\
\email{seok@kias.re.kr} }

\author{Kimyeong Lee\\
\\
School of Physics, Korea Institute for Advanced Study, Seoul 130-012 KOREA\\
\email{klee@kias.re.kr} }

\abstract{We find new $\frac{1}{8}$-BPS giant graviton solutions
in $AdS_5 \times S^5$, carrying three angular momenta along $S^5$,
and investigate their properties. Especially, we show that nonzero
worldvolume gauge fields are admitted preserving supersymmetry.
These gauge field modes can be viewed as electromagnetic waves
along the compact D3 brane, whose Poynting vector contributes to
the BPS angular momenta. We also analyze the (nearly-)spherical
giant gravitons with worldvolume gauge fields in detail.
Expressing the $S^3$ in Hopf fibration ($S^1$ fibred over $S^2$),
the wave propagates along the $S^1$ fiber.}


\begin{document}

\section{Introduction and Conclusion}

1/2 BPS gravitons with high angular momentum in $AdS_p\times S^q$
are shown to be expanded into higher dimensional brane objects,
the so-called giant gravitons, by McGreevy, Susskind and
Toumbas\cite{mst}. The massless gravitons get polarized to branes
due to the Myers dielectric effect\cite{myers}. These giant
gravitons can be BPS objects\cite{gmt,hhi}, and also have been
studied in relation to the AdS-CFT correspondence.

Less supersymmetric, 1/4 and 1/8 BPS, giant gravitons have also
been found. Giant gravitons can be considered as finite D-branes,
which can be studied by the worldvolume theory defined by the
Dirac-Born-Infeld(DBI) and Chern-Simons(CS) actions. Especially
supersymmetric giant gravitons extended in $S^q$ were
characterized by the intersection of $S^q$ and a holomorphic
surface by Mikhailov\cite{mikh}. See also \cite{ha-smi,acr}. Giant
graviton as an extended D-brane could in principle carry dynamical
world-volume gauge field, still keeping supersymmetry.
\cite{sa-sh,pr-sh} treat some related objects, corresponding to
strings ending on/dissolved into giant gravitons. In this work we
show that this is possible and find some exact properties of the
BPS gauge field living on the giant graviton of the Type IIB
string theory on $AdS_5\times S^5$, which contributes to both
energy and angular momentum on $S^5$. Especially we find the
explicit 1/8 BPS smooth electromagnetic wave solutions on a
nearly-spherical BPS D3 brane, and study their quantum physics.
(We will not treat the dual giant gravitons\cite{gmt,hhi} in this
paper.)

While the general electromagnetic wave propagating on a three
manifold would be very complicated, we find that the
supersymmetric one is considerably simpler.  This solution of
course satisfies Gauss-Bianchi constraints. Especially when the
world volume geometry of a giant graviton is $S^3$, which can be
regarded as a Hopf-bundle of $S^1$ over $S^2$, we find explicit
smooth configuration for all BPS electromagnetic waves.  While the
bundle structure is what makes such wave possible, roughly one can
see that electric and magnetic fields are mutually orthogonal but
of the same magnitude on $S^2$ and  propagate along the $S^1$
fibre, so that the Poynting vector density is along the $S^1$
direction at each point. Such configuration turns out to be smooth
and has finite energy. (As for the previous work for the study of
fluctuations around spherical giant gravitons, see \cite{djm}.)
When a 1/8 BPS giant graviton is moving along $S^5$ with three
angular momentum, such an electromagnetic wave on world volume
seems to be still possible, preserving the same 1/8
supersymmetries.

For our giant gravitons, conserved quantities are three angular
momenta, say, $J_{12}$, $J_{34}$, $J_{56}$ along the $S^5$. Thus,
such 1/8 BPS electromagnetic wave contributes to the BPS energy
and angular momentum, which leads to additional degeneracy of the
1/8 giant graviton quantum states.  Here we  quantize  all BPS
electromagnetic waves on $S^3$, leading to quantized angular
momentum contributions.

One immediate question is whether one can obtain more explicit
solutions for the gauge fields when the shape of giant graviton is
more complicated. In this paper we constructed explicit solutions
for the nearly-spherical case. Presumably, 1/8 BPS giant gravitons
can have more complicated topology, like three torus and so on,
which may allow also 1/8 BPS electromagnetic wave. (Topologically
nontrivial giant gravitons are constructed in the maximally
supersymmetric plane wave background of M-theory\cite{mikh2,bkl}.)
In addition, there could be also nontrivial gauge holonomy and/or
flux along non-contractible cycle. It would be interesting to find
such solutions explicitly. In somewhat different direction, there
is some work on giant gravitons with nonzero gauge fields on the
plane wave background obtained from the Penrose limit of
$AdS_5\times S^5$ \cite{taka}. Our work could be generalized to
the plane wave case and shed some light on the subject.

Quantum mechanically, there is an enormous degeneracy of giant
gravitons with given angular momenta, which could be countable in
principle. When gravitational back reaction is included, such BPS
object in AdS space does not appear as an extremal black hole, but
as a `superstar' with null or time-like naked
singularity\cite{ta-my}. See also \cite{surya}. Recently, regular
solutions of the 10 dimensional supergravity with one angular
momentum has been studied\cite{llm}, and there also has been some
study of non-supersymmetric black holes carrying more than one
charges\cite{gu-he}. However, the complete understanding of the
quantum degeneracy of giant graviton states and the counting of
these seems to be somewhat wanting.

A new class of extremal black hole solutions have been found in
$AdS$ space\cite{gut-rea}. Besides angular momentum along $S^5$,
they carry angular momentum also in the $AdS_5$ part. This
solution has singularities cloaked inside a horizon with nonzero
area. Thus it would be interesting to find BPS giant gravitons,
with worldvolume electromagnetic wave, which carries angular
momenta in the $AdS_5$ part also. $\frac{1}{2}$-BPS solutions of
this type exist in the reference \cite{adkss,ca-si}.

The gauge field solutions found in this paper prompt us to
consider the $\frac{1}{8}$-BPS giant gravitons in $AdS_4\times
S^7$ made of $M5$ branes with four angular momenta\cite{mikh},
including the self-dual three form tensor field strength on the
worldvolume. It is naturally conceivable that one needs to
consider $S^5$ as $S^1$ fibration over ${\mathbb{CP}}^2$
\cite{kk-sk}. In a related maximally supersymmetric plane wave
background, BPS tensor modes around $\frac{1}{2}$-BPS vacuum have
been observed \cite{msr}. It would be desirable to have clear
geometric understanding as we got through the work of this paper.

In the matrix theory context, interesting observations have been
made through a series of papers by Janssen \textit{et.al.}
\cite{jlr3,jlr5}, which we think is somewhat related to our
present work as well as future projects. They constructed
$\frac{1}{2}$-BPS spherical giant gravitons in the $AdS_5\times
S^5$ and $AdS_4\times S^7$ cases from the relevant matrix
theories. The $S^1$ fibrations over suitable projective spaces are
considered (taking advantage of fuzzy ${\mathbb{CP}}^1\!=\!S^1$
and ${\mathbb{CP}}^2$) to form $S^3$ and $S^5$ giant gravitons.

The organization of this paper is as follows. In section 2 we
review the construction of $\frac{1}{8}$-BPS giant gravitons
without turning on world volume gauge fields. In section 3 we show
that gauge fields can be turned on in a supersymmetric way. First
we provide the general condition for the gauge fields to preserve
$\frac{1}{8}$ supersymmetry. Then we consider the Gauss law and
related constraints which should be further satisfied. We also
compute the energy and angular momenta on $S^5$ carried by these
configurations, and show that energy saturates the BPS bound given
by sum of three angular momenta. In section 4 we provide the exact
solutions of the constraint equations for gauge field on a
spherical giant graviton and show that they are electromagnetic
waves propagating along closed circles in Hopf fibration of $S^3$.
We also quantize this explicit solution, assuming small
fluctuations, and identify the angular momentum quanta of these
modes. One appendix is included to explain technical facts.

\section{$\frac{1}{8}$-BPS giant gravitons from holomorphic surfaces}

In this section we review the giant graviton solutions without
turning on worldvolume gauge fields. We will also clarify our
notations and conventions.


\vskip 0.6cm\hspace{-0.6cm}{\bf\large 2.1\ \ \ Supersymmetry of
$AdS_5\times S^5$ background}\vskip 0.3cm

\hspace{-0.6cm}We start by embedding $AdS_5\times S^5$ with radii
$R$ into a 12 dimensional space
$\mathbb{R}^{2+4}\times\mathbb{R}^6$, which would be useful
throughout this paper. Writing the two radial coordinates of
$\mathbb{R}^{2+4}$ and $\mathbb{R}^{6}$ as $r_1$ and $r_2$,
respectively, the metric on $AdS_5\times S^5$ is inherited from
that of the flat space in a manifest way:
\begin{equation}
  ds_{\mathbb{R}^{2+4}}^2=-R^2 dr_1^2+r_1^2ds_{AdS_5}^2\ ,\ \
  ds_{\mathbb{R}^6}^2=R^2 dr_2^2+r_2^2ds_{S^5}^2
\end{equation}
with restriction to $r_1=r_2=1$ subspace. The spin connection
components containing $r_1$ or $r_2$ indices are as follows
(characters with caret are local orthonormal frame indices):
\begin{equation}\label{12 connection}
  \omega^{\hat{\mu}\hat{r}_1}=-\frac{1}{R}\ e^{\hat{\mu}}\ ,\ \
  \omega^{\hat{i}\hat{r}_2}=\frac{1}{R}\ e^{\hat{i}}
\end{equation}
where $\mu$ and $i$ denote five $AdS_5$ and $S^5$ indices,
respectively, in appropriate coordinates, and $e^{\hat{a}}$ is the
vielbein 1-form.

We summarize the construction of 32 Killing spinors in
$AdS_5\times S^5$ with self-dual 5-form fluxes, starting from 12
dimensional covariantly constant spinors. The Killing spinors
should leave the IIB gravitino invariant under the following
supersymmetry transformation
\begin{equation}\label{gravitino}
  \delta\psi_M=\frac{1}{\kappa}D_M\epsilon+
  \frac{i}{4\kappa\cdot480}F^{(5)}_{NPQRS}\Gamma^{NPQRS}\Gamma_M\epsilon=0
\end{equation}
where $D_M\epsilon=\partial_M\epsilon+\frac{1}{4}\omega_{M}^{\
\hat{P}\hat{Q}}\Gamma_{\hat{P}\hat{Q}}\epsilon$. The IIB chirality
condition is
\begin{equation}\label{IIB chirality}
  \Gamma^{\hat{0}\hat{1}\hat{2}\hat{3}\hat{4}}\Gamma^{
  \hat{5}\hat{6}\hat{7}\hat{8}\hat{9}}\epsilon=+\epsilon\ ,
\end{equation}
where $0\sim 4$/$5\sim 9$ are $AdS_5$/$S^5$ indices, respectively.
The flux part is given as
\begin{equation}\label{5 form flux}
  \frac{1}{480}F^{(5)}_{NPQRS}\Gamma^{NPQRS}=\frac{1}{R}
  (\Gamma^{\hat{5}\hat{6}\hat{7}\hat{8}\hat{9}}+
  \Gamma^{\hat{0}\hat{1}\hat{2}\hat{3}\hat{4}})\
  \overset{eff}{=}\frac{2}{R}\Gamma^{\hat{5}\hat{6}\hat{7}\hat{8}\hat{9}}
\end{equation}
when acting on antichiral spinors, like $\Gamma_{\!M}\epsilon$ in
(\ref{gravitino}).

To solve this Killing spinor equation, we start from a 12
dimensional Dirac spinor $\Psi$, which has $2^{6}=64$ complex
components. We will assume the Majorana representation with real
gamma matrices. We require it to be covariantly constant in 12
dimensional sense. In the most trivial frame for the 12
dimensional vielbein, $\Psi$ is simply a constant spinor since
$\mathbb{R}^{2+4}\times\mathbb{R}^6$ is flat. However, performing
a local Lorentz transformation to make
$\Gamma^{\hat{0}}\sim\Gamma^{\hat{9}}$ and $\Gamma^{\hat{r}_1}$,
$\Gamma^{\hat{r}_2}$ as numerical matrices, $\Psi$ gains
nontrivial dependence on the $AdS_5\times S^5$ coordinates (but
not on $r_1$ or $r_2$). Let us consider $\Psi$ in this frame with
the following two projection constraints (they are numerical
projectors in both frames)
\begin{equation}\label{6D projection}
  \Gamma^{\hat{0}\hat{1}\hat{2}\hat{3}\hat{4}\hat{r}_1}\Psi=-i\Psi\ ,\ \
  \Gamma^{\hat{5}\hat{6}\hat{7}\hat{8}\hat{9}\hat{r}_2}\Psi=+i\Psi\ .
\end{equation}
Our convention for gamma matrices is
$(\Gamma^{\hat{0}})^2=(\Gamma^{\hat{r}_1})^2=-1$, while all the
others square to 1. Then, using the expression (\ref{12
connection}), the 12 dimensional covariant constancy condition is
rephrased in terms of $AdS_5\times S^5$ coordinates as
\begin{eqnarray}
  0&=&D_{\mu}\Psi+\frac{1}{2}\omega_{\mu}^{\ \hat{\nu}\hat{r}_1}
  \Gamma_{\hat{\nu}\hat{r}_1}\Psi=D_\mu\Psi+
  \frac{i}{2R}\Gamma^{\hat{0}\hat{1}\hat{2}\hat{3}\hat{4}}\Gamma_{\mu}\Psi\nonumber\\
  0&=&D_{i}\Psi+\frac{1}{2}\omega_{i}^{\ \hat{j}\hat{r}_2}
  \Gamma_{\hat{j}\hat{r}_2}\Psi=D_i\Psi+
  \frac{i}{2R}\Gamma^{\hat{5}\hat{6}\hat{7}\hat{8}\hat{9}}\Gamma_{i}\Psi
  \label{killing}
\end{eqnarray}
which is nearly, but not exactly yet, the Killing spinor equation
(\ref{gravitino}) in $AdS_5\times S^5$ with the flux (\ref{5 form
flux}). To complete the construction, we have to make sure that
IIB chirality condition (\ref{IIB chirality}) is satisfied. $\Psi$
does not satisfy this condition, but the projected spinor
\begin{equation}\label{projection}
  \epsilon\equiv\frac{1-\Gamma^{\hat{r}_1\hat{r}_2}}{2}\Psi\ \ \ ,
  \ \ \ \ \Gamma^{\hat{r}_1\hat{r}_2}\epsilon=-\epsilon
\end{equation}
does. Since this projector commutes with all the matrices
appearing in (\ref{killing}), we can make $\epsilon$ satisfy the
same equation. However, after replacing $\Psi$ by $\epsilon$,
$\Gamma^{\hat{0}\hat{1}\hat{2}\hat{3}\hat{4}}$ in (\ref{killing})
can be replaced by $\Gamma^{\hat{5}\hat{6}\hat{7}\hat{8}\hat{9}}$,
and we finally get the desired Killing spinor equation. As a 12
dimensional spinor, the final answer $\epsilon$ is subject to two
projection conditions (\ref{IIB chirality}) and
(\ref{projection}). Therefore, it carries 16 complex components,
as required for the $AdS_5\times S^5$ Killing spinor.


\vskip 0.6cm\hspace{-0.6cm}{\bf\large 2.2\ \ \ $\frac{1}{8}$-BPS
giant gravitons}\vskip 0.3cm

\hspace{-0.6cm}In this subsection we review the D3 giant gravitons
preserving $\frac{1}{8}$ supersymmetry and carrying three
components of $SO(6)$ angular momentum, using holomorphic
surfaces\cite{mikh}. We will present the details since it will be
useful in the next section.

The D3 brane we are interested in stays at the origin of $AdS_5$
and has nontrivial shape and time evolution in $S^5$. It is
constructed by the following procedure:
\begin{enumerate}

\item Regarding the embedding space $\mathbb{R}^6$ as
$\mathbb{C}^3$ with holomorphic coordinates $Z_1$, $Z_2$, $Z_3$,
consider any holomorphic surface given by an equation of the form
$F(Z_1,Z_2,Z_3)=0$.

\item Let us call $\Sigma$ the 3-manifold given by the
intersection of the above surface and the sphere
$|Z_1|^2+|Z_1|^2+|Z_1|^2=1$. This is the D3 brane configuration at
a given time, say $t=0$, where $t$ is the worldvolume time
coordinate. We partially fix the worldvolume diffeomorphism by
setting $Rt$ to be the proper time for an observer sitting at the
origin of $AdS_5$.

\item Given the above initial configuration, the time evolution of
the brane is given as follows. The coordinate $Z_k$ of a point on
the 3-manifold evolves as $\dot{Z}_k=iZ_k$, where dot denotes $t$
derivative. The worldvolume trajectory of the D3 brane is thus
given as $F(e^{-it}Z_1,e^{-it}Z_2,e^{-it}Z_3)=0$.

\end{enumerate}
In the rest of this subsection we will summarize the proof that
this configuration preserves $\frac{1}{8}$ supersymmetry.

To simplify the proof, let us follow \cite{mikh} and introduce
some useful notations. Let ${\bf e}^{\perp}$ be a unit vector in
$\mathbb{R}^6$ which is normal to $S^5$. The time evolution
$\dot{Z}_k=iZ_k$ can be phrased in a different way that the
velocity vector is $I.\ {\bf e}^\perp$ at each point of D3 brane,
where the operation $I$ gives the complex structure to
$\mathbb{R}^6$.\footnote{Writing the components as $Z_k=X_k+iY_k$,
and corresponding unit vectors as $\hat{x}_k$ and $\hat{y}_k$, we
have $I.\hat{x}_k=\hat{y}_k$ and $I.\hat{y}_k=-\hat{x}_k$.}
However, the vector $I.\ {\bf e}^\perp$ is not orthogonal to the
spatial D3 manifold $\Sigma$, so it is not the physical velocity.
We introduce another unit vector ${\bf e}^\phi\in T(S^5)$, aligned
toward the direction of transverse velocity. Another unit vector
in $T(S^5)$ transverse to $\Sigma$ and normal to ${\bf e}^\phi$ is
called ${\bf e}^n$. One can easily see that
\begin{equation}\label{linear dependent}
  I.\ {\bf e}^\phi=-\cos\alpha\ {\bf e}^\perp+\sin\alpha\ {\bf e}^n\ ,
\end{equation}
and that $\cos\alpha={\bf e}^\phi\cdot I.\ {\bf e}^\perp\equiv v$
is the transverse velocity of the D3 brane in suitable worldvolume
orientation. See appendix A for the proof.

The supersymmetry preserved by above configuration can be checked
by investigating the supersymmetry plus compensating kappa
symmetry transformation. For the D3 brane, it is given as follows
(we follow the notation of the second reference in \cite{be-to}):
\begin{equation}\label{kappa projection}
  \Gamma\epsilon=\epsilon
\end{equation}
where $\epsilon$ is the Killing spinor obtained in the previous
subsection, and
\begin{eqnarray}\label{kappa definition}
  &&\Gamma=\frac{1}{\sqrt{\det(1+Y)}}\left[
  i\sigma_2\otimes\left(1+\frac{1}{8}\gamma^{\mu\nu\rho\sigma}F_{\mu\nu}
  F_{\rho\sigma}\right)\Gamma_{(0)}-\sigma_1
  \otimes\left(\frac{1}{2}\gamma^{\mu\nu}F_{\mu\nu}
  \right)\Gamma_{(0)}\right]\nonumber\\
  &&Y^{\mu}_{\ \ \nu}\equiv g^{\mu\rho}F_{\rho\nu}\ \ \ ,\ \ \ \
  \Gamma_{(0)}\equiv\frac{1}{4!\sqrt{-\det g}}\ \epsilon^{\mu\nu\rho\sigma}
  \gamma_{\mu\nu\rho\sigma}\ \ \ \ ((\Gamma_{(0)})^2=-1)\\
  &&\epsilon_{0123}=-\epsilon^{0123}=+1\ \ \ ,\ \ \ \
  \gamma_{\mu}=\Gamma_{i}\frac{\partial
  X^i}{\partial\sigma^{\mu}}\ :\ {\rm\ induced\ gamma\ matrix}\ .\nonumber
\end{eqnarray}
The $2\times 2$ Pauli matrices act on the $SL(2,\mathbb{R})$
indices of the type IIB spinors. Since we are using the complex
Killing spinor, they act as \cite{imam}
\begin{equation}
  i\sigma_2 \epsilon =-i\epsilon\ \ , \ \ \ \sigma_1 \epsilon=
  i\epsilon^\ast\ \ ,\ \ \ \sigma_3 \epsilon=\epsilon^\ast\ .
\end{equation}
We will consider the inclusion of gauge field $F_{\mu\nu}$ in the
next section.

Without the gauge fields, the projection operator becomes
\begin{equation}
  \Gamma=-i\Gamma_{(0)}=\frac{-i}{\sqrt{1-v^2}}
  (\Gamma^{\hat{0}}-v\Gamma^{\hat{\phi}})\hat{\Sigma}
\end{equation}
where $\Gamma^{\hat\phi}=\Gamma({\bf e}^\phi)$ is the Gamma matrix
along ${\bf e}^\phi$ direction which squares to 1, and
$\hat{\Sigma}$ is the product of three Gamma matrices
corresponding to the spatial part of D3 worldvolume satisfying
$\hat{\Sigma}^2=-1$. Note that $\Gamma^{\hat{\phi}}$ and
$\hat{\Sigma}$ anticommute since ${\bf e}^\phi$ is transverse to
the D3 brane. Using the identification $v=\cos\alpha$ and the
relation (\ref{linear dependent}), the supersymmetry condition
(\ref{kappa projection}) can be written as
\begin{equation}\label{no gauge projection}
  0=(1-\Gamma^{\hat{r}_1\hat{r}_2})\Gamma^{\hat{0}\hat{r}_1 \hat{n}\hat{\phi}}
  \left[1-\Gamma^{\hat{0}\hat{r}_1}
  \Gamma({\bf e}^{\phi})\Gamma(I.\ {\bf e}^\phi)\right]\Psi\ \ \rightarrow\ \
  \Gamma^{\hat{0}\hat{r}_1}\Gamma({\bf e}^{\phi})\Gamma(I.\ {\bf e}^\phi)\Psi=+\Psi\
  .
\end{equation}
In obtaining this condition, we used the orientation convention
\begin{equation}\label{R6 orientation}
  \hat{\Sigma}\Gamma^{\hat{n}\hat{\phi}\hat{r}_2}\Psi=
  \Gamma^{\hat{5}\hat{6}\hat{7}\hat{8}\hat{9}\hat{r}_2}=i\Psi
\end{equation}
together with the second condition of (\ref{6D projection}). We
will work in the trivial frame where $\Psi$ is a constant spinor.
The solution for (\ref{no gauge projection}) in generic case is
obtained as follows. First, (\ref{no gauge projection}) is solved
by imposing\footnote{Of course the other sign choice also solves
(\ref{no gauge projection}). However, it turns out that only one
of these two is compatible with (\ref{6D projection}).}
\begin{equation}\label{decompose projection}
  \Gamma^{\hat{0}\hat{r}_1}\Psi=+i\Psi\ \ ,\ \ \
  \Gamma({\bf e}^\phi)\Gamma(I.\ {\bf e}^\phi)\Psi=-i\Psi\ .
\end{equation}
Note that the first projector is a numerical matrix also in the
trivial frame, since the D3 brane is sitting at the origin of
$AdS_5$. Writing the latter projector as
\begin{equation}
  \Gamma\left(\frac{{\bf e}^\phi+iI.\ {\bf e}^\phi}{2}\right)
  \Gamma\left(\frac{{\bf e}^\phi-iI.\ {\bf e}^\phi}{2}\right)
  =\frac{1-i\Gamma({\bf e}^\phi)\Gamma(I.\ {\bf e}^\phi)}{2}\ ,
\end{equation}
we are led to the nontrivial requirement
\begin{equation}\label{holo projection 1}
  \Gamma\left(\frac{{\bf e}^\phi-iI.\ {\bf
  e}^\phi}{2}\right)\Psi=0\ .
\end{equation}
Since this matrix generically depends on all the $S^5$
coordinates, the only way to fulfill the requirement is to set
\begin{equation}\label{holo projection 2}
  \Gamma(\partial_{Z_k})\Psi=
  \frac{1}{2}\left(\frac{}{}\Gamma_{X_k}\!-\!i\Gamma_{Y_k}\right)\Psi=0
\end{equation}
for all pair indices $k=1,2,3$. These three projections are not
independent: one is given by the other two using the condition
(\ref{6D projection}). Therefore, two of these three together with
the first of (\ref{decompose projection}) make this configuration
$\frac{1}{8}$-BPS.

\section{Giant gravitons with worldvolume gauge fields}

Having reviewed the giant gravitons without worldvolume gauge
fields, we now turn to the generalization with gauge fields turned
on.


\vskip 0.6cm\hspace{-0.6cm}{\bf\large 3.1\ \ \
Supersymmetry}\vskip 0.3cm

\hspace{-0.6cm}We first set our notation for the induced metric on
the worldvolume. It can be written as
\begin{equation}\label{wv metric}
  g_{\mu\nu}=R^2\left(
  \begin{array}{cc}-1+v^2&0\\0&h_{ij}\end{array}
  \right)\ \ , \ \ \ h_{ij}=e^{\hat{k}}_{i}e^{\hat{k}}_{j}\ ,
\end{equation}
where $i,j,k=1,2,3$. The absence of $g_{0i}$ components may be
understood as being killed by time-dependent diffeomorphism. The
quantity $e^{\hat{i}}_j$ is the spatial vielbein on D3. Its
inverse is written as $e^{i}_{\hat{j}}$, satisfying
$e^{\hat{i}}_{k}\ e^{k}_{\hat{j}}=\delta^{\hat{i}}_{\hat{j}}$ ,
etc. The vielbein should not be confused either with the bulk
vielbein used in the previous section, or with the various unit
vectors written in bold characters. We are also going to specify a
convenient expression for the spatial worldvolume metric $h_{ij}$
in (\ref{wv metric}). On the 3 manifold $\Sigma$ given by
holomorphic surface as in the previous section, there exists an
$I$-invariant sub-plane $T_0\Sigma$ in the tangent space at each
point: let us call the unit vector (normalized by induced metric
on $\Sigma$) normal to this plane as ${\bf e}^\psi$, following
\cite{mikh}. One can easily see from the definition that
\begin{equation}\label{velocity decomp}
  I.{\bf e}^\perp=\sin\alpha\ {\bf e}^\psi+
  \cos\alpha\ {\bf e}^\phi\ .
\end{equation}
At a given moment of time (say $t=0$), we can choose one of our
spatial coordinate as $\psi$ such that its associated tangent
vector $\partial_{\psi}$ is proportional to ${\bf e}^\psi$. Then,
one can write the general metric as follows:
\begin{equation}\label{wv spatial metric}
  ds_\Sigma^2=h(x,\psi)\left(d\psi+\sum_{a=1,2}V_a(x,\psi) dx^a\right)^2+
  g(x,\psi)\sum_{a=1,2}(dx^a)^2\ .
\end{equation}
We took advantage of $x^1$-$x^2$ diffeomorphism to go to a sort of
conformal gauge and have a common factor $g(x,\psi)$. The vielbein
components are given as follows:
\begin{eqnarray}
  &&e^{\hat{\psi}}=\sqrt{h}(d\psi+V_a dx^a)\ \ ,\ \ \
  e^{\hat{a}}=\sqrt{g}dx^a\nonumber\\
  &&e_{\hat{\psi}}=\frac{1}{\sqrt{h}}\partial_\psi\ \ ,\ \ \
  e_{\hat{a}}=\frac{1}{\sqrt{g}}\left(\partial_{a}-V_a\partial_\psi\right)\ ,\label{wv vielbein}
\end{eqnarray}
which should not be confused with the bulk vielbein we used in
section 2. The choice of inverse vielbein $e_{\hat{\psi}}$ is
indeed proportional to $\partial_\psi$, as we required.

The electric and magnetic fields are defined as
\begin{equation}
  E_{\hat{i}}=\frac{1}{\sqrt{1-v^2}}f^{i}_{\hat{j}}F_{0j}\ \ ,\ \ \
  B_{\hat{i}}=\frac{1}{2}\epsilon_{ijk}f^{l}_{\hat{j}}f^{m}_{\hat{k}}
  F_{lm}\ \ \ (\epsilon_{123}=1)\ .
\end{equation}
The square root factor is introduced for convenience. We will use
the rescaled field strength $F_{\mu\nu}=R^2 F_{\mu\nu}^{(scaled)}$
in order not to have the $R^2$ factors here and there. With this
convention, we compute various $F_{\mu\nu}$-dependent quantities
appearing in (\ref{kappa definition}). The relevant quantities are
expressed as
\begin{eqnarray}
  \det(1+Y)&=&1+|\vec{B}|^2-|\vec{E}|^2-(\vec{E}\cdot\vec{B})^2\ \ ,\nonumber\\
  \frac{1}{8}\gamma^{\mu\nu\rho\sigma}F_{\mu\nu}F_{\rho\sigma}&=&
  (\vec{E}\cdot\vec{B})\ \Gamma_{(0)}\ \ ,\\
  \frac{1}{2}\gamma^{\mu\nu}F_{\mu\nu}=\gamma^{\hat{t}\hat{i}}E_{\hat{i}}+
  \frac{1}{2}\epsilon_{ijk}\gamma^{\hat{i}\hat{j}}B_{\hat{k}}&=&
  \frac{1}{\sqrt{1-v^2}}(\Gamma^{\hat{0}}-v\Gamma^{\hat{\phi}})(\vec{\gamma}\cdot\vec{E})+
  \hat{\Sigma}(\vec{\gamma}\cdot\vec{B})\ .\nonumber
\end{eqnarray}
where the vectors denote spatial 3-vectors with indices expressed
in local orthonormal frame on the worldvolume, and we used
$\epsilon_{\hat{1}\hat{2}\hat{\psi}}=1$. We also note that the
sign convention for the tensor $\epsilon_{ijk}$ on $\Sigma$ should
be $\epsilon^{\hat{x}\hat{y}\hat{\psi}}\!=\!1$, where $x,y$ are
the indices parametrizing the 2 manifold and become `x-like' and
`y-like' variables, respectively, after being push-forwarded: that
is, $I.e_{\hat{1}}=e_{\hat{2}}$. This can be shown by a careful
sign check using the convention (\ref{R6 orientation}) and the $I$
operation rules (\ref{linear dependent}), (\ref{velocity decomp}).

Since our main motivation is looking for the states preserving the
\textit{same supersymmetry}, we require the relation
\begin{equation}\label{same susy}
  -i\Gamma_{(0)}\epsilon=\epsilon\ \ \rightarrow\ \
  \frac{1}{\sqrt{1-v^2}}(\Gamma^{\hat{0}}-v\Gamma^{\hat{\phi}})\epsilon=
  i\hat{\Sigma}\epsilon\ ,
\end{equation}
or equivalently, (\ref{holo projection 2}) that we developed in
the previous section. Therefore, we still have the same solution
for the shape and its time evolution given by holomorphic
surfaces. With (\ref{same susy}) assumed, the supersymmetry
condition (\ref{kappa projection}) becomes
\begin{equation}\label{kappa with gauge}
  \frac{1}{\sqrt{1+|\vec{B}|^2-|\vec{E}|^2-(\vec{E}\cdot\vec{B})^2}}
  \left[1+i(\vec{E}\cdot\vec{B})-(\sigma_1)\otimes\hat{\Sigma}
  \vec{\gamma}\cdot(\vec{E}+i\vec{B})
  \right]\epsilon=\epsilon\ ,
\end{equation}
where $\sigma_1$ is understood to act on the whole complex
quantity $(\vec{E}+i\vec{B})\epsilon$.

First of all, since the action of $\sigma_1$ is complex
conjugation on $\epsilon$ (or $(\vec{E}+i\vec{B})\epsilon$), it is
hard to expect supersymmetry if this term does not
vanish.\footnote{At the end of this subsection, we will show this
term should vanish indeed to have supersymmetry.} Therefore we
require
\begin{equation}\label{gauge 1}
  \vec{\gamma}\cdot(\vec{E}+i\vec{B})\epsilon=0\ .
\end{equation}
Then, looking at the remaining terms in (\ref{kappa with gauge}),
we should also require
\begin{equation}\label{gauge 2}
  \vec{E}\cdot\vec{B}=0\ \ ,\ \ \ |\vec{E}|=|\vec{B}|
\end{equation}
to have supersymmetry.

We now present some useful facts to solve (\ref{gauge 1}) and
(\ref{gauge 2}) using the $\frac{1}{8}$ supersymmetry condition
(\ref{holo projection 2}). At each point $x$ on $\Sigma$, recall
that there is a vector field ${\bf e}^\psi$ which does not close
to $T\Sigma$ under the action of $I$, and a two dimensional
subspace $T_0\Sigma$ orthogonal to ${\bf e}^\psi$ which is closed
under $I$ \cite{mikh}. With our coordinate and vielbein choice
(\ref{wv spatial metric}) and (\ref{wv vielbein}), one obtains
\begin{equation}
  \gamma_{\hat{1}}-i\gamma_{\hat{2}}=\Gamma^{A}
  \frac{\partial X^A}{\partial\sigma^{k}}
  \ (e^{k}_{\hat{1}}-ie^{k}_{\hat{2}})\equiv
  \Gamma\left(e_{\hat{1}}-ie_{\hat{2}}\right)
\end{equation}
where $e_{\hat{a}}$ ($a=1,2$) are understood as push-forwards of
the worldvolume vectors $e_{\hat{a}}^i$. Since $e_{\hat{a}}$ are
the two orthonormal vectors spanning $T_0\Sigma$, we have
\begin{equation}\label{wv holo gamma}
  (e_{\hat{2}})^A=
  \left(I.\ e_{\hat{1}}\right)^A\ \
  \rightarrow\ \
  \gamma_{\hat{1}}-i\gamma_{\hat{2}}=
  \Gamma\left(e_{\hat{1}}-iI.e_{\hat{1}}\right)\ .
\end{equation}
We used the fact that vectors $e_{\hat{1}}$ and $e_{\hat{2}}$
behaves respectively as `x' and `y' direction after being
push-forwarded, and not vice versa, as mentioned above. From
(\ref{holo projection 1}) and (\ref{holo projection 2}), we
finally observe that (\ref{wv holo gamma}) implies
$(\gamma_{\hat{1}}-i\gamma_{\hat{2}})\Psi=0$.

The first requirement (\ref{gauge 1}) can be written as
\begin{equation}
  (1-\Gamma^{\hat{r}_1\hat{r}_2})
  \left[\gamma^{\hat{a}}E^{\hat{a}}+
  i(\epsilon_{ab}\gamma^{\hat{b}})(\epsilon_{ac}B^{\hat{c}})+
  \gamma^{\hat{\psi}}(E^{\hat{\psi}}+iB^{\hat{\psi}})\right]\Psi=0\
  \ \ (a=1,2,\ \epsilon_{12}=1)\ .
\end{equation}
The first two terms can annihilate $\Psi$ by choosing
$B^{\hat{a}}=\epsilon_{ab}E^{\hat{b}}$, which can be seen from
$(\gamma_{\hat{1}}-i\gamma_{\hat{2}})\Psi=0$. The last term has to
be zero by itself, which requires
$E^{\hat{\psi}}=B^{\hat{\psi}}=0$. They can be summarized by a
single equation
\begin{equation}\label{gauge requirement}
  \vec{B}=-I.\vec{E}\ \ ,
\end{equation}
where the vectors are understood to be push-forwarded. Then the
second requirement (\ref{gauge 2}) is also satisfied. Even after
this restriction, we have two real functions as remaining degrees:
the magnitude $|\vec{E}|=|\vec{B}|$ and the overall rotation
degree of these vectors on the $I$ invariant plane.

We finally comment that the requirement (\ref{gauge 1}) is indeed
the most general one in the supersymmetry class (\ref{holo
projection 2}). First, one can easily check directly from
(\ref{kappa with gauge}) that
$B^{\hat{a}}=\epsilon_{ab}E^{\hat{b}}$ has to be imposed:
otherwise there cannot be any supersymmetry due to the appearance
of matrices like $\gamma^{\hat{\psi}\hat{a}}$. Then, one may keep
nonzero $E_{\psi}$ and $B_{\psi}$ together with
$B_{\hat{a}}=\epsilon_{ab}E_{\hat{b}}$ to solve the supersymmetry
condition (\ref{kappa with gauge}) directly. The resulting
condition is
\begin{equation}
  \frac{1}{\sqrt{(1+B_{\hat{\psi}}^{\ 2})(1-E_{\hat{\psi}}^{\
  2})}}
  \left(\begin{array}{cc}
  1+E_{\hat{\psi}}&-B_{\hat{\psi}}(1+E_{\hat{\psi}})\\
  -B_{\hat{\psi}}(1-E_{\hat{\psi}})&1-E_{\hat{\psi}}
  \end{array}\right)
  \left(\begin{array}{c}R\\I\end{array}\right)=
  \left(\begin{array}{c}R\\I\end{array}\right)
\end{equation}
where $R$/$I$ denotes the real/imaginary part of the 12
dimensional spinor $\Psi\!=\!R+iI$, respectively, in Majorana
representation (spinor indices suppressed). This eigenvector
equation can be satisfied only when
$E_{\hat{\psi}}=B_{\hat{\psi}}=0$.


\vskip 0.6cm\hspace{-0.6cm}{\bf\large 3.2\ \ \ Local (and global)
constraints}\vskip 0.3cm

\hspace{-0.6cm}Apart from the supersymmetry requirement
(\ref{gauge requirement}), we also have to impose the Gauss law
constraint and the Bianchi identity: the latter has to be checked
also since we have not expressed field strengths in terms of
vector potential. To check the Gauss law, we have to compute the
electric displacement. The DBI Lagrangian is
\begin{equation}
   {\mathcal{L}}_{DBI}=-R^4\sqrt{\det h}\sqrt{1-v^2}\sqrt{
   1+|\vec{B}|^2-|\vec{E}|^2-(\vec{E}\cdot\vec{B})^2}\ ,
\end{equation}
and the Chern-Simon term would not give any contribution. The
electric displacement, after imposing the condition (\ref{gauge
2}), becomes
\begin{equation}
  \Pi^i=\frac{\partial{\mathcal{L}}}{\partial F_{0i}}=
  \frac{1}{\sqrt{1-v^2}}\ e^{i}_{\hat{j}}\
  \frac{\partial{\mathcal{L}}}{\partial E_{\hat{j}}}=
  R^4 \sqrt{\det h}\ e^{i}_{\hat{j}} E^{\hat{j}}\ .
\end{equation}
The Gauss constraint is
\begin{equation}\label{gauss}
  \partial_i \Pi^i=0\ \ \rightarrow\ \
  \partial_i\left(\sqrt{\det h}\ e^{i}_{\hat{j}}
  E^{\hat{j}}\right)=0\ .
\end{equation}
The Bianchi identities become
\begin{eqnarray}
  \partial_{[i}F_{jk]}=0 &\rightarrow&
  \partial_i\left(\sqrt{\det h}\ e^{i}_{\hat{j}}
  B^{\hat{j}}\right)=0\ ,\label{bianchi 1}\\
  \partial_0 F_{ij}+\partial_{i}F_{j0}+\partial_{j}F_{0i}=0
  &\rightarrow&
  \partial_t\left(\sqrt{\det h}\ e^{i}_{\hat{j}}B^{\hat{j}}\right)=
  \epsilon^{ijk}\partial_{[j}\left(\sqrt{1\!-\!v^2}e^{\hat{l}}_{k]}E_{\hat{l}}\right)\
  .  \label{bianchi 2}
\end{eqnarray}
The second Bianchi identity (\ref{bianchi 2}) tells us the time
evolution of $\vec{B}$ once it is given at initial time. We will
not regard it as a constraint: it will be treated as providing
time evolution of $\vec{B}$ in the next section.

Here we consider the constraints (\ref{gauss}) and (\ref{bianchi
1}) at given time. We plug the vielbein (\ref{wv vielbein}) and
$E^{\hat{\psi}}=0$ into the two constraints (\ref{gauss}) and
(\ref{bianchi 1}) to obtain
\begin{eqnarray}
  &&\partial_{a}\left( \sqrt{hg}\ E^{\hat{a}}\right)=
  \partial_\psi\left(V_a \sqrt{hg}\ E^{\hat{a}}\right)\nonumber\\
  &&\partial_{a}\left( \sqrt{hg}\ B^{\hat{a}}\right)=
  \partial_\psi\left(V_a \sqrt{hg}\ B^{\hat{a}}\right)\ ,\ \ (a=1,2\ {\rm summed})
\end{eqnarray}
where the caret indices are again the local orthonormal frame
ones. Combining the two coordinates $x^1$ and $x^2$ into
\begin{equation}
  z\equiv x^1+ix^2\ ,\ \ \bar{z}\equiv z^\ast\ \ \rightarrow\ \
  \partial\equiv\frac{\partial}{\partial z}=
  \frac{1}{2}\left(\partial_{1}-i\partial_{2}\right)\ ,\  \
  \bar{\partial}\equiv\frac{\partial}{\partial \bar{z}}=
  \frac{1}{2}\left(\partial_{1}+i\partial_{2}\right)
\end{equation}
the above two constraints are written as
\begin{eqnarray}
  &&\partial\left(\sqrt{hg}\ \bar{E}\right)+
  \bar{\partial}\left(\sqrt{hg}\ E\right)=
  \frac{1}{2}\left\{\partial_\psi (\bar{V}\sqrt{hg}\ E)+
  \partial_\psi(V \sqrt{hg}\ \bar{E})\right\}\ \
  ,\nonumber\\
  &&\partial\left(\sqrt{hg}\ \bar{B}\right)+
  \bar{\partial}\left(\sqrt{hg}\ B\right)=
  \frac{1}{2}\left\{\partial_\psi (\bar{V}\sqrt{hg}\ B)+
  \partial_\psi(V \sqrt{hg}\ \bar{B})\right\}\ \
  ,\label{constraints}
\end{eqnarray}
where $E\equiv E^{\hat{1}}-iE^{\hat{2}}$, $B\equiv
B^{\hat{1}}-iB^{\hat{2}}$ and $V\equiv V_{1}-iV_{2}$. The
supersymmetry requirement (\ref{gauge requirement}) can be
reexpressed as $B=iE$, which allows us to write
(\ref{constraints}) as a single complex equation
\begin{equation}\label{constraint holo}
  \bar{\partial}\left(\sqrt{hg}\ E\right)=
  \frac{1}{2}\ \partial_\psi
  \left(\bar{V}\sqrt{hg}\ E\right)\ .
\end{equation}
For the general metric of the form (\ref{wv metric}), it does not
look easy to get an explicit solution. Here we will obtain the
formal solution of this constraint, but we will also present an
explicit solution for the nearly-spherical case in the next
section.

We expand the functions $\sqrt{hg}\ E$ and $V$ appearing in
(\ref{constraint holo}) as $\partial_\psi$ eigenmodes, i.e.,
\begin{eqnarray}
  \sqrt{hg}\ E(z,\bar{z},\psi)&=&\sum_{n=-\infty}^\infty
  (\sqrt{hg}\ E)_n(z,\bar{z})\ e^{-in\psi}\ ,\nonumber\\
  V(z,\bar{z},\psi)&=&\sum_{n=-\infty}^\infty
  V_n(z,\bar{z})\ e^{-in\psi}\ ,\label{constraint sol 1}
\end{eqnarray}
where $n$ runs over a suitable multiple of integers, depending on
the $\psi$ period. Then the constraint (\ref{constraint holo}) is
written as
\begin{equation}
  \bar{\partial}\left(\sqrt{hg}\ E\right)_n=
  -\frac{in}{2}\sum_{m=-\infty}^{\infty}(\bar{V})_{n-m}(\sqrt{hg}\ E)_m\ .
\end{equation}
The formal solution for this equation is
\begin{equation}\label{formal sol}
  \left(\sqrt{hg}\ E\right)_n=\sum_{m=-\infty}^{\infty}P\exp\left(-\frac{in}{2}\int
  d\bar{z}\ \bar{V}\right)_{nm}\ G_m(z)
\end{equation}
where $\bar{V}$ is an $\infty\times\infty$ matrix with entry
$\bar{V}_{mn}=\bar{V}_{m-n}$, and the expression `$P\exp$'
(together with an integral $\int d\bar{z}$) denotes the standard
path-ordered product of matrices.

The above formal expression looks messy and not so illuminating.
Here we simplify this formal solution for a special case where
$\bar{V}$ becomes independent of $\psi$ coordinate. This
simplified form will be used to obtain an explicit solution in the
next section.\footnote{Currently, the only example for this
$\psi$-independent $\bar{V}$ we know is the spherical giant
graviton, which will be considered in the next section.} In this
setting, we only need to consider the mode expansion of
$\sqrt{hg}\ E$ in (\ref{constraint sol 1}). Inserting this
expansion into (\ref{formal sol}), we get the decoupled expression
for each modes
\begin{equation}\label{constraint sol 2}
 (\sqrt{hg}\ E)_n(z,\bar{z})=G_n(z)
 \exp\left(-i\frac{n}{2}\int d\bar{z}\ \bar{V}(z,\bar{z})\right)\ .
\end{equation}
The integration in the exponent is an indefinite integral. The
holomorphic functions $G_n(z)$ are the integration constants,
which are the arbitrary functions surviving the (local)
constraints. Note that the $\sqrt{hg}E$ is invariant under the
coordinate transformation $\psi\rightarrow \psi+ f(z,\bar{z}) $
compensated by a transformation of $V$ which leaves the metric
invariant when $h,V,g$ are independent of $\psi$.

Note that all the analysis so far does not take any global issues
into account. Since the coordinates $x^1$ and $x^2$ parametrize a
compact 2 manifold, they may develop coordinate singularities at
certain points. We should require the solution (\ref{constraint
sol 1}) and (\ref{constraint sol 2}) to be well-behaved at these
points. In the next section we will give a concrete illustration
how to take care of this global constraint, with nearly-spherical
giant gravitons as a simple example. Here we present general
expectation.

At coordinate singularities (\ref{constraint sol 2}) may be
divergent: divergent solutions are accompanied with unwanted
singular sources for the left hand sides of (\ref{constraints}).
We require the function $G_n(z)$ to be sufficiently regular near
such coordinate singularities, so as to tame the potential
singularities in (\ref{constraint sol 2}) and leave
(\ref{constraints}) source-free. Suppose we chose the coordinate
such that there is a coordinate singularity at $z=0$. Then,
discarding the singular modes would truncate the Laurent expansion
of $G_n(z)$ into a sort of Taylor expansion. When the 2 manifold
has the topology of $S^2$, as we will study in the next section,
there are two coordinate singularities. Two such truncations
should be imposed in this case. It would not always be true that
there are terms surviving both truncations: there may or may not
exist such terms depending on the sign of $n$ in the exponent of
(\ref{constraint sol 2}).

There is another form of regularity requirement for $E$: the
energy carried by the gauge field has be finite. We will compute
the energy for our BPS configuration in the next subsection, but
this criterion should be related to the above source-free
condition. We will consider both constraints with the
nearly-spherical giant graviton in the next section.


\vskip 0.6cm\hspace{-0.6cm}{\bf\large 3.3\ \ \ Energy and angular
momenta of the giant graviton}\vskip 0.3cm

\hspace{-0.6cm}In this subsection, we compute the gauge field
contribution to the energy and sum of three $SO(6)$ angular
momenta $J_k\equiv J_{x_k y_k}$ ($k=1,2,3$). These two quantities
turn out to be same, showing that the energy carried by the gauge
field modes saturates the BPS bound given by the sum of angular
momenta.

We first compute the canonical energy. It is given by
\begin{equation}\label{energy}
  {\mathcal{E}}=\dot{\vec{X}}\cdot\frac{\partial{\mathcal{L}}}{\partial\dot{\vec{X}}}+
  F_{0i}\frac{\partial{\mathcal{L}}}{\partial
  F_{0i}}-{\mathcal{L}}=
  R^4\sqrt{\det h_{ij}}\ \frac{1+|\vec{E}|^2}{\sqrt{1-v^2}}
\end{equation}
after using the supersymmetry conditions. Note that, using the
supersymmetry condition $\vec{B}=-I.\vec{E}$, the term depending
on gauge field may also be written as
$|\vec{E}|^2=|\vec{E}\times\vec{B}|$.

To calculate the angular momenta, we compute the canonical momenta
conjugate to the coordinates $\vec{X}$, which may be regarded as
living in ${\mathbb{R}}^6$. These momenta can be divided into two
parts: those coming from the DBI action and from Chern-Simons
term. To compute the DBI contribution of the canonical momenta
\begin{equation}
  \vec{P}_{DBI}=\frac{\partial{\mathcal{L}}_{DBI}}{\partial\dot{\vec{X}}}
  =-\frac{1}{2}\sqrt{-\det(g+F)_{\mu\nu}}\ [(g+F)^{-1}]^{\mu\nu}
  \frac{\partial}{\partial\dot{\vec{X}}}\ [g+F]_{\mu\nu} \ ,
\end{equation}
we should first do the $\dot{\vec{X}}$ derivative without fixing
the worldvolume gauge like (\ref{wv metric}), and set
$\dot{\vec{X}}\cdot\partial_i\vec{X}=0$ ($i=1,2,3$) afterward. In
the notation of previous subsections, the relevant quantities are
given as follows (after imposing the supersymmetry condition
(\ref{gauge requirement})):
\begin{eqnarray}
  \sqrt{-\det(g+F)_{\mu\nu}}&=&R^4\sqrt{\det
  h}\sqrt{1-v^2}\ \ ,\nonumber\\
  \left[(g+F)^{-1}\right]^{00}&=&-\frac{1}{R^2}\frac{1+|\vec{B}|^2}{1-v^2}\ \ ,\\
  \left[(g+F)^{-1}\right]^{(0i)}&=&\frac{1}{R^2\sqrt{1-v^2}}\ e^{i}_{\hat{j}}\
  (\vec{E}\times\vec{B})^{\hat{j}}\ \ ,\nonumber\\
  \frac{\partial}{\partial\dot{\vec{X}}}\ [g+F]_{00}=2R^2 \vec{v}\
  &,&\ \frac{\partial}{\partial\dot{\vec{X}}}\ [g+F]_{0 i}=R^2\partial_i\vec{X}
  \ \ , \nonumber
\end{eqnarray}
where the parenthesis on indices means symmetrization. Therefore,
we obtain
\begin{equation}\label{DBI mom}
  \vec{P}_{DBI}=R^4\sqrt{\det h}\left[
  (1+|\vec{B}|^2)\frac{\vec{v}}{\sqrt{1-v^2}}\ -\ \partial_i\vec{X}\
  e^{i}_{\hat{j}}(\vec{E}\times\vec{B})^{\hat{j}}
  \right]\ .
\end{equation}
The first term is transverse to the 3 manifold $\Sigma$, while the
second term is longitudinal. Looking at this second term, the
vector $\vec{E}\times\vec{B}$ has $\hat\psi$ component only.
Furthermore, from,
\begin{equation}
  \vec{X}_i\ e^i_{\hat\psi}={\bf e}^\psi\ ,
\end{equation}
this longitudinal term is simplified to be
\begin{equation}
  -\ \partial_i\vec{X}\
  e^{i}_{\hat{j}}(\vec{E}\times\vec{B})^{\hat{j}}
  =-{\bf e}^\psi\ (\vec{E}\times\vec{B})^{\hat\psi}\ \ .
\end{equation}
The cross product $(\vec{E}\times\vec{B})^{\hat\psi}$ in the
second term is simply $-|\vec{B}|^2$, from the supersymmetry
requirement (\ref{gauge requirement}) and the worldvolume
orientation $\epsilon_{\psi 12}=1$ that we chose.

What we need is the sum of three angular momenta, an $SO(6)$
generator corresponding to the rotation $\delta\vec{X}\propto
I.\vec{X}$. This is nothing but the velocity vector, decomposed
into transverse and longitudinal parts as (\ref{velocity decomp}).
One may rewrite it a bit differently as
\begin{equation}
  I.\vec{X}=\vec{v}\ +\sqrt{1-v^2}\ {\bf e}^\psi\ .
\end{equation}
The DBI contribution to the sum of three angular momenta is given
as
\begin{equation}\label{DBI angular}
  [J_1+J_2+J_3]_{DBI}=(I.\vec{X})\cdot\vec{P}_{DBI}=
  R^4\sqrt{\det h}\ \
  \frac{v^2+|\vec{B}|^2}{\sqrt{1-v^2}}\ ,
\end{equation}
which is not the same as the energy (\ref{energy}) yet. The
contribution from the Chern-Simons term to the sum of three
angular momenta is computed in \cite{mikh}, with the solutions
without worldvolume gauge fields. We can use that result since, in
our background, Chern-Simons term is unchanged after turning on
gauge fields. The result is
\begin{equation}\label{CS angular}
  [J_1+J_2+J_3]_{CS}=R^4\sqrt{\det h}\sqrt{1-v^2}\ .
\end{equation}
Adding (\ref{DBI angular}) and (\ref{CS angular}), we obtain
\begin{equation}
  J_{12}+J_{34}+J_{56}=R^4\sqrt{\det h}\ \
  \frac{1+|\vec{B}|^2}{\sqrt{1-v^2}}\ ,
\end{equation}
which is exactly the energy (\ref{energy}). Thus we have checked
that the energy of giant graviton saturates the BPS bound given by
three $SO(6)$ charges even after the gauge fields are turned on.

\section{(Nearly-)spherical solutions and quantization}

In this section we explicitly construct the gauge field solutions.
We will consider nearly-spherical giant gravitons. We will also
quantize these modes when the fluctuation is small.

\vskip 0.6cm\hspace{-0.6cm}{\bf\large 4.1\ \ \ The explicit
solution\vskip 0.3cm}

\hspace{-0.6cm}The holomorphic surface for a nearly spherical
giant graviton having large angular momentum in the $X_3$-$Y_3$
plane is given by the equation
\begin{equation}\label{near surface}
  Z_3= Z_3^{(0)}+f(Z_1,Z_2)\ ,\ \ |Z_1|^2+|Z_2|^2+|Z_3|^2=1\ ,
\end{equation}
where $f$ is a holomorphic function, much smaller than
$Z_3^{(0)}$. We also fix the worldvolume diffeomorphism on
$\Sigma$ using a natural parametrization of $S^3$: we choose the
three coordinates $\alpha,\phi_1,\phi_2$ on $S^3$ (thus on
$\Sigma$) as follows:
\begin{eqnarray}\label{sphere coordi}
  &&Z_1=\sin\Theta_{0}\cos\alpha e^{i\phi_1}\ ,\ \
  Z_2=\sin\Theta_{0}\sin\alpha e^{i\phi_2}\ \ \ \
  (Z_3^{(0)}\equiv\cos\Theta_{0}\ e^{i\phi_{0}}=
  v\ e^{i\phi_{0}})\nonumber\\
  &&ds^2_{S^3}=d\alpha^2+\cos^2\alpha\ d\phi_1^2+\sin^2\alpha\
  d\phi_2^2\ .
\end{eqnarray}
The ranges of the variables are given as
$0\leq\alpha\leq\frac{\pi}{2}$ and $\phi_1\sim\phi_1+2\pi$,
$\phi_2\sim\phi_2+2\pi$.

We first identify the induced vector field $\sin\alpha\ {\bf
e}^\psi$, along the direction of which the gauge fields should
vanish. It can be easily obtained from (\ref{velocity decomp}) as
\begin{equation}\label{sphere isometry}
  \sin\alpha\ {\bf e}^\psi=I.\ {\bf e}^\perp-\ {\bf n}_1\
  \frac{{\bf n}_1\cdot I.\ {\bf e}^\perp}{{\bf n}_1\cdot\ {\bf n}_1}
  -\ {\bf n}_2\
  \frac{{\bf n}_2\cdot I.\ {\bf e}^\perp}{{\bf n}_2\cdot\ {\bf n}_2}
  \approx
  (iZ_1,iZ_2,0)+{\mathcal{O}}(f)
\end{equation}
where ${\bf n}_1$ and ${\bf n}_2$ are two vectors normal to the
holomorphic surface $F(Z_1,Z_2,Z_3)=0$:
\begin{equation}
  {\bf n}_1=(f_1, f_2,-1)\ ,\ \ {\bf n}_2=i(f_1,f_2,-1)\ ,
  \ \ {\bf n}_1\cdot\ {\bf n}_2=0.
\end{equation}
Here we used the target ${\mathbb{C}}^3$ indices like
$(Z_1,Z_2,Z_3)$ for the vectors, which are related to the
${\mathbb{R}}^6$ indices like $Z_k=X_k+iY_k$. To the leading order
in $f$, the vector field (\ref{sphere isometry}) is
$\partial_{\phi_1}+\partial_{\phi_2}$ in our coordinate system
(\ref{sphere coordi}), and this should be proportional to the
vector $\partial_\psi$ in the metric (\ref{wv spatial metric}).
After doing the following coordinate transformation
\begin{equation}
  \psi\equiv\phi_1+\phi_2\ ,\ \ \phi\equiv\phi_1-\phi_2\ ,\ \
  \theta\equiv 2\alpha\ ,
\end{equation}
we get the $S^3$ metric in Hopf fibration
\begin{equation}\label{hopf}
  4 ds^2_{S^3}=d\theta^2+\sin^2\theta d\phi^2 +(d\psi+\cos\theta
  d\phi)^2
\end{equation}
with the coordinate range given as
\begin{equation}
  0\leq\theta\leq\pi\ ,\ \phi\sim\phi+2\pi\ ,\
  \psi\sim\psi+4\pi\ .
\end{equation}
Therefore, the base 2 manifold is $S^2$, having two coordinate
singularities at $\theta=0,\pi$, as mentioned in the previous
section.

One can easily obtain the holomorphic coordinates and relevant
complex functions from the metric (\ref{hopf}):
\begin{equation}\label{sphere functions}
  z=2\tan\left(\frac{\theta}{2}\right)e^{i\phi}\ ,\ \
  h=\frac{1}{4}(1-v^2)\ ,\ \
  g=\frac{1}{4}(1-v^2)\left(1+\frac{z\bar{z}}{4}\right)^{-2}\ ,\ \
  \bar{V}=\frac{i}{\bar{z}}\ \frac{4-z\bar{z}}{4+z\bar{z}}\ .
\end{equation}
Since $\psi$ is $4\pi$-periodic, $n$ in (\ref{constraint sol 1})
assumes half integer values. The solution for $(\sqrt{hg}\ E)_n$,
given by (\ref{constraint sol 2}), is calculated to be
\begin{eqnarray}
  \exp\left\{-i\frac{n}{2}\int d\bar{z}\bar{V}(z,\bar{z})\right\}
  G_n(z)&=&\left[\frac{z\bar{z}}{(4+z\bar{z})^2}\right]^{n/2}G_n(z)
  \nonumber\\
  &\sim&\left(\sin\frac{\theta}{2}\right)^{n}
  \left(\cos\frac{\theta}{2}\right)^{n}
  G_n(z)\ ,\label{reg constraint}
\end{eqnarray}
which should be sufficiently regular near $\theta\rightarrow
0,\pi$, respectively, in order not to have singular sources there.
Let us make a Laurent expansion of $G_{n}(z)$:
\begin{equation}
  G_{n}(z)=\sum_{k=-\infty}^{\infty}\frac{a^\ast_{n,k}}{z^{k}}\ ,
\end{equation}
where we included the complex conjugation for the coefficients
$a_{n,k}$ for later convenience. The requirement for the $2\pi$
periodicity of $\phi_1$ and $\phi_2$ is that $n$ and $k$ should be
either integers or half the odd integers at the same time.
Furthermore, forbidding singular sources at $\theta=0, \pi$, one
gets the condition
\begin{equation}
  {\rm Regularity\ at\ }
  \left\{\begin{array}{l}
  \theta=0:\ n-k\geq 0\\
  \theta=\pi:\ n+k\geq 2
  \end{array}\right\}
  \rightarrow\ -n+2\leq k\leq n.
\end{equation}
The total number of modes for given $\psi$-momentum $n$ is
$2n\!-\!1$, and the allowed values for n are $1$, $\frac{3}{2}$,
$2$, $\frac{5}{2}, \cdots$. Especially, there are no
$\psi$-independent modes, i.e., $n=0$. This is natural since the
Gauss-Bianchi constraint (\ref{constraint holo}) would reduce to
that on $S^2$-base for $n=0$, which looks too restrictive to admit
regular solutions.

To write down the mode expansion, it is more illuminating to
advocate a sort of polar basis for the complex fields $E$ and $B$,
given as follows:
\begin{equation}
  E^{polar}\equiv E^{\hat{\theta}}-iE^{\hat{\phi}}=
  e^{i\phi}(E^{\hat{1}}-iE^{\hat{2}})=e^{i(\phi_1-\phi_2)}E^{Cart.}
\end{equation}
where we included the superscript `$Cart.$' to emphasize that
complex field we used so far is in Cartesian basis. In this polar
basis, we have the neat expression for the mode expansion given as
\begin{equation}
  \sin\theta\ E^{polar}(\theta,\phi_1,\phi_2)=
  \sum_{l_1,l_2=1}^{\infty}a^\ast_{l_1 l_2}\
  e^{-il_1\phi_1}e^{-il_2\phi_2}\left(\cos\frac{\theta}{2}\right)^{l_1}
  \left(\sin\frac{\theta}{2}\right)^{l_2}
\end{equation}
where $l_1\equiv n+k-1$ and $l_2\equiv n-k+1$ runs over
$1,2,3,\cdots$, and $a_{l_1 l_2}$'s are complex numbers. This
expression will turn out to be the most natural one in the next
subsection, in that the angular momenta $J_{1}$ and $J_{2}$ along
the $Z_1$ and $Z_2$ plane would be $l_1$ and $l_2$, respectively,
for each mode.

Note that there is no mode with either of $l_1$ and $l_2$ being
zero. This is in contrast to the mechanical fluctuation which
contains the modes with either of the two angular momenta being
zero.\footnote{This can be checked straightforwardly by using the
holomorphic surface solutions, keeping the leading contribution of
$f$ in (\ref{near surface}).} The electromagnetic fields fall to
zero at $\theta=0,\pi$ (in the orthonomal frame units with caret
indices) except for the lowest mode $l_1=l_2=1$ (or $n=1$)
\begin{equation}
  E^{polar}_{1,1}(\theta,\phi_1,\phi_2)=-iB^{polar}_{1,1}(\theta,\phi_1,\phi_2)=
  a^\ast_{1,1}\ e^{-i\phi_1}e^{-i\phi_2}\ .
\end{equation}
Together with $\partial_\psi$, these lowest $E_{1,1}^{polar}$ and
$B_{1,1}^{polar}$ form a threesome of nowhere-vanishing
orthonormal vector fields on $S^3$.

Finally, let us check the time evolution of these modes. It is
given by the second Bianchi `identity' (\ref{bianchi 2}) and the
remaining equation of motion
\begin{equation}\label{eom}
  \partial_{\mu}\left(\frac{\partial{\mathcal{L}}}{\partial F_{\mu
  i}}\right)=0\ \ \rightarrow\
  \partial_t\left(\sqrt{\det h}\ e^{i}_{\hat{j}}E^{\hat{j}}\right)=
  -\epsilon^{ijk}\partial_{[j}\left(\sqrt{1\!-\!v^2}\ e^{\hat{l}}_{k]}B_{\hat{l}}\right)\
  .
\end{equation}
In general there should be more subtlety since the spatial
coordinate frames we have chosen may change by time evolution,
thus requiring additional terms due to compensating gauge
transformation. However, it does not matter in our spherical case.
In this case (\ref{sphere functions}), the two equations
(\ref{bianchi 2}) and (\ref{eom}) are combined into one
holomorphic equation and a real $\psi$-component equation:
\begin{eqnarray}
  \partial_t E&=&2i\partial_\psi B=-2\partial_\psi
  E\nonumber\\
  \partial_t(\sqrt{g}V_aE^{\hat{a}})&=&2\epsilon_{ab}
  \partial_a\left(\sqrt{g}B_{\hat{b}}\right)
  =-2\partial_a\left(\sqrt{g}E_{\hat{a}}\right)
\end{eqnarray}
where we used $B=iE$ (or $B_{\hat{a}}=\epsilon_{ab}E_{\hat{b}}$)
to replace all $B$'s into $E$'s. Then, expressing every field
strengths and $V$ with their holomorphic components, we get
\begin{eqnarray}
  \partial_t E&=&-\ 2\partial_{\psi}E\\
  \partial_t (\bar{V}E+V\bar{E})&=&-
  2\left(\bar{V}\partial_\psi E +V\partial_\psi \bar{E}
  \right)\nonumber
\end{eqnarray}
where we popped out $\sqrt{g}$ or $V_{a}$'s from $\partial_t$
since they are all time-independent for the spherical giant case.
Inserting the mode expansion $E(z,\bar{z},\psi)=\sum_n
E_{n}(z,\bar{z})e^{-in\psi}$, all these equations are solved by
giving the time evolution
\begin{eqnarray}
  \sin\theta\ E^{polar}(z,\bar{z},\psi,t)&=&\sum_{n=1}^\infty
  \sin\theta\ E_{n}^{polar}(z,\bar{z})e^{-in(\psi-2t)}\nonumber\\
  &=&\!\!\sum_{l_1,l_2=1}^\infty
  a^\ast_{l_1 l_2}\left(\cos\frac{\theta}{2}\right)^{l_1}\!\!
  \left(\sin\frac{\theta}{2}\right)^{l_2}\!
  e^{-il_1(\phi_1-t)}e^{-il_2(\phi_2-t)}.\label{spherical modes}
\end{eqnarray}
Note that, the phase velocity
$\psi/t=2=\sqrt{\frac{-g_{tt}}{g_{\psi\psi}}}$ is the light
velocity along the $\psi$ direction, measured by the worldvolume
metric (\ref{wv metric}) and (\ref{sphere functions}).


\vskip 0.6cm\hspace{-0.6cm}{\bf\large 4.2\ \ \ Quantizing the
small fluctuations}\vskip 0.3cm

\hspace{-0.6cm}In this subsection we consider small fluctuation of
gauge fields on a (nearly-)spherical giant graviton and quantized
them. We will also show that each mode carries integer-valued
angular momenta given by $l_1$ and $l_2$ identified in
(\ref{spherical modes}).\footnote{The quantization of mechanical
BPS fluctuation can also be done, using nearly-spherical
holomorphic surfaces, and following the procedure of
\cite{cp-mar,bho,bhko}. We will not show this result here.}

We quantize the BPS gauge field modes identified in the previous
section assuming the fluctuation is `small' : the meaning of the
latter will be addressed more quantitatively as we proceed. To do
so, we first compute the DBI Lagrangian up to quadratic order in
the gauge field strength $F_{\mu\nu}$ (or, equivalently, $\vec{E}$
and $\vec{B}$):
\begin{equation}
  {\mathcal{L}}_{DBI}\cong R^4\sqrt{\det{h}}\sqrt{1\!-\!v^2}
  \left(-1+\frac{1}{2}|\vec{E}|^2-\frac{1}{2}|\vec{B}|^2+h.o.t.\right)\
  .
\end{equation}
This expansion is valid as long as $|\vec{E}|^2\!\ll\!1$. This
condition will finally be translated into the smallness of
occupation numbers after quantization. What we would like to keep
is the quadratic term proportional to $|\vec{E}|^2$, which is the
kinetic term and should tell us the structure of quantization. We
choose the temporal gauge
\begin{equation}
  A_0=0\ \rightarrow\ F_{0i}=\dot{A}_{i}\ (i=1,2,3)\ .
\end{equation}
Since we are only interested in quantizing the BPS modes, we take
advantage of the fact $F_{0\psi}=0$ and hence set $A_{\psi}=0$.
The quadratic kinetic term can be rewritten in the first order
form as follows:
\begin{eqnarray}
  \frac{R^4}{2}\sqrt{\det{h}}\sqrt{1\!-\!v^2}|\vec{E}|^2&=&
  \frac{1}{2}\left(R^4\sqrt{\det{h}}f^{i}_{\hat{j}}E^{\hat{j}}\right)
  \left(\sqrt{1\!-\!v^2}f_i^{\hat{k}}E_{\hat{k}}\right)\nonumber\\
  &=&\Pi^i\dot{A}_i-\frac{\sqrt{1\!-\!v^2}}{2R^4\sqrt{\det{h}}}g_{ij}
  \Pi^i\Pi^j\ .
\end{eqnarray}
We will try to do the mode expansion of the first term with the
coefficient $a_{l_1 l_2}$ defined in (\ref{spherical modes}),
regarded as off-shell degrees of freedom.

To this end, let us first express the vector potential $A_i$ with
this BPS mode expansion. First we re-express the on-shell mode
expansion of $E$ as
\begin{equation}\label{t deri mode}
  \sin\theta\ E^{polar}=\frac{\partial}{\partial t}\left[\sum_{l_1,l_2=1}^{\infty}
  \frac{a^\ast_{l_1 l_2}}{i(l_1+l_2)}e^{i(l_1+l_2)t}
  \left(\cos\frac{\theta}{2}\right)^{l_1}\left(\sin\frac{\theta}{2}\right)^{l_2}
  e^{-il_1\phi_1}e^{-il_2\phi_2}\right]\ .
\end{equation}
Recalling the definition of the vector potential, one gets
\begin{equation}
  \sin\theta\ E^{polar}=e^{i\phi}\frac{\sin\theta}{\sqrt{1\!-\!v^2}}
  \left(f^1_{\hat{1}}\dot{A}_1-if^2_{\hat{2}}\dot{A}_2\right)=
  \frac{\sin\theta}{\sqrt{1\!-\!v^2}\sqrt{g}}\ \dot{A}^{polar}
\end{equation}
where we used the complexified vector potential
$A^{polar}=e^{i\phi}A\equiv e^{i\phi}(A_1\!-\!iA_2)$. The series
expansion in the square bracket of (\ref{t deri mode}) is
essentially the expression for $A^{polar}$, but let us absorb the
on-shell time evolution factor $e^{i(l_1+l_2)t}$ into $a^\ast_{l_1
l_2}$ as time-dependent $a^\ast_{l_1 l_2}(t)$ and pretend the
off-shell expression
\begin{equation}
  A^{polar}(t)=\frac{\sqrt{1\!-\!v^2}\sqrt{g}}
  {\sin\theta}\sum_{l_1,l_2=1}^{\infty}\frac{a^\ast_{l_1
  l_2}(t)}{i(l_1+l_2)}
  \left(\cos\frac{\theta}{2}\right)^{l_1}\left(\sin\frac{\theta}{2}\right)^{l_2}
  e^{-il_1\phi_1}e^{-il_2\phi_2}\ .
\end{equation}
One should use this off-shell expression since we are going to
read off the quantization rule from the Lagrangian defined with
the off-shell fields. The term $\Pi^i\dot{A}_i$ containing the
information on canonical structure can be re-written as
\begin{equation}
  \frac{R^4}{2}\sqrt{hg}\ \bar{E}\dot{A}+ (h.c.)=
  \frac{R^4}{2}\sqrt{hg}\ \bar{E}^{polar}\dot{A}^{polar}+ (h.c.)\ ,
\end{equation}
where $E$ and $E^{polar}$ are also understood as off-shell
expressions, replacing $a^\ast_{l_1 l_2}e^{i(l_1+l_2)t}$ by
$a^\ast_{l_1 l_2}(t)$. To get the mode expansion of the action, we
should also do the integration
\begin{equation}\label{ang integration}
  \int_0^{4\pi}d\psi\int dx^1 dx^2 =
  \int_0^\pi d\theta\int_0^{2\pi}d\phi_1\int_0^{2\pi} d\phi_2\
  4\sec^2\frac{\theta}{2}\ \tan\frac{\theta}{2}\ .
\end{equation}
After the integration, the mode expansion for the action is
\begin{equation}\label{canonical exp}
  \frac{i}{2}\sum_{l_1, l_2=1}^\infty\
  \frac{\pi^2 R^4}{2}(1-v^2)^2\frac{(l_1-1)!(l_2-1)!}{(l_1+l_2)!}\ a_{l_1
  l_2}^\ast\ \dot{a}_{l_1 l_2}\ +\ (h.c.)\ +\ \cdots\ .
\end{equation}
Therefore, after being promoted to operators, the modes $a_{l_1
l_2}$ satisfy the harmonic oscillator commutation relation in
suitable normalization:
\begin{equation}
  \left[a_{l_1 l_2},a_{m_1 m_2}^\dag\right]=
  \frac{2}{\pi R^4(1-v^2)^2}\ \frac{(l_1+l_2)!}{(l_1-1)!(l_2-1)!}
  \ \delta_{l_1, m_1}\delta_{l_2, m_2}\ .
\end{equation}
The number operators
\begin{equation}
  N_{l_1 l_2}=\frac{\pi R^4(1-v^2)^2}{2}\frac{(l_1-1)!(l_2-1)!}{(l_1+l_2)!}
  \ a_{l_1 l_2}^\dag a_{l_1 l_2}
\end{equation}
assume integer eigenvalues.

At this point we turn to the question of the `smallness' of
fluctuations, i.e., try to rephrase the criterion
$|\vec{E}|^2\!\ll\!1$ in quantum language. Regarded as an operator
with mode expansion (\ref{spherical modes}), and also with
worldvolume integration, the condition $\int\sqrt{\det h}\
|\vec{E}|^2 \ll\int\sqrt{\det h}$ is given as
\begin{equation}\label{small condition}
  \sum_{l_1,l_2=1}^{\infty}(l_1+l_2)N_{l_1l_2}\ll R^4(1-v^2)^2\ .
\end{equation}

We finally compute the quantized angular momentum operators $J_1$
and $J_2$'s along $X_1$-$Y_1$ and $X_2$-$Y_2$ planes in terms of
the number operators. They are given by
\begin{equation}
 J_1=X^1 [P_{DBI}]_{Y_1}-Y^1 [P_{DBI}]_{X_1}\ ,\ \
 J_2=X^2 [P_{DBI}]_{Y_2}-Y^2 [P_{DBI}]_{X_2}\ ,
\end{equation}
where $P_{DBI}$ is given as (\ref{DBI mom}). Inserting
\begin{equation}
  (X^1,Y^1)=\sqrt{1\!-\!v^2}\cos\frac{\theta}{2}(\cos\phi_1,\sin\phi_1)\
  ,\ (X^2,Y^2)=\sqrt{1\!-\!v^2}\sin\frac{\theta}{2}(\cos\phi_2,\sin\phi_2),
\end{equation}
we obtain the following expressions for the angular momentum
densities
\begin{equation}
  J_1=R^4\sqrt{1\!-\!v^2}\sqrt{hg^2}\cos^2\frac{\theta}{2}|\vec{E}|^2\ ,\ \
  J_2=R^4\sqrt{1\!-\!v^2}\sqrt{hg^2}\sin^2\frac{\theta}{2}|\vec{E}|^2\
  .
\end{equation}After inserting
(\ref{sphere functions}), (\ref{spherical modes}) and doing the
worldvolume integration, the angular momentum operators become
\begin{equation}
  J_1=\sum_{l_1,l_2}l_1\hat{N}_{l_1 l_2}\ \ \ ,\ \ \
  J_2=\sum_{l_1,l_2}l_2\hat{N}_{l_1 l_2}\ .
\end{equation}
The third angular momentum $J_3$ along the $X_3$-$Y_3$ plane,
which should be much larger than the other two in our nearly
spherical setting, also carries nonzero contribution from the
gauge modes. It is given by
\begin{equation}\label{macro ang mom}
  J_3=2\pi^2 R^4(1-v^2)+\frac{v^2}{1-v^2}\sum_{l_1 l_2=1}^\infty(l_1+l_2)
  \hat{N}_{l_1 l_2}
\end{equation}
where the first and second terms are the contributions from the
mechanical part and gauge field fluctuations, respectively. The
second term is always much smaller than the first mechanical
contribution, taking (\ref{small condition}) and $v<1$ into
account.

\vskip 1cm

{\bf\Large\hspace{-0.8cm} Acknowledgements}

\hspace{-0.6cm}K.L. is supported in part by KOSEF
R01-2003-000-10319-0.

\vskip 1cm

{\bf\Large\hspace{-0.8cm} Appendix}



\renewcommand{\thesection}{\Alph{section}}

\setcounter{section}{1} \setcounter{equation}{0}

\vskip 0.6cm\hspace{-0.6cm}{\bf\Large A\ \ \ Properties of various
vector fields}\vskip 0.3cm

\hspace{-0.6cm}In this appendix, we briefly recover the properties
of various vector fields ${\bf e}^\perp$, ${\bf e}^\phi$, ${\bf
e}^\psi$. It is essentially repeating \cite{mikh}, to clarify our
convention. We start by a unit vector ${\bf e}^\perp=(X_k,Y_k)$
perpendicular to $S^5$. The velocity vector $I.\ {\bf e}^\perp$ is
decomposed into transverse and longitudinal components with
respect to the 3 manifold $\Sigma$, with the unit vector fields
${\bf e}^\phi$ and ${\bf e}^\psi$ defined only on $\Sigma$. There
is another vector in $TS^5$ orthogonal to both $\Sigma$ and ${\bf
e}^\phi$, which we call ${\bf e}^n$. The action of $I$ on these
vectors are given as follows:
\begin{eqnarray}
  I.\ {\bf e}^\perp&=&\cos\alpha\ {\bf e}^\phi+\sin\alpha\ {\bf e}^\psi\label{rot1}\\
  I.\ {\bf e}^n&=&-\sin\alpha\ {\bf e}^\phi+ \cos\alpha\ {\bf e}^\psi\label{rot2}\\
  I.\ {\bf e}^\phi&=&-\cos\alpha\ {\bf e}^\perp+\sin\alpha\ {\bf e}^n\label{rot3}\\
  I.\ {\bf e}^\psi&=&-\sin\alpha\ {\bf e}^\perp -\cos\alpha\ {\bf e}^n\ \label{rot4}.
\end{eqnarray}
The first one just follows from the definition of ${\bf e}^\psi$
and $T_0\Sigma$. We set the velocity $v=\cos\alpha$ to be
positive, and also set $\sin\alpha$ to be positive, which is the
convention for ${\bf e}^\psi$. We will derive the remaining three
relations, trying to distinguish the derived facts and
conventions.

Recall that there is a sub-plane $T_0\Sigma$ of $T\Sigma$ which is
invariant under the action of $I$. In fact, such pair of
directions in $\Sigma$ is guaranteed to exist from the fact that
it is constructed from the intersection of a holomorphic surface
and $S^5$. Since ${\bf e}^\perp$ is orthogonal to $T_0\Sigma$, so
is $I.\ {\bf e}^\perp$. Otherwise $T_0\Sigma$ would not be
invariant under $I$. Therefore, ${\bf e}^\psi$ is orthogonal to
$T_0\Sigma$ from its definition (\ref{rot1}). We have four unit
vector fields ${\bf e}^\perp$, ${\bf e}^\phi$, ${\bf e}^n$ and
${\bf e}^\psi$ orthogonal to $T_0\Sigma$, which should also close
under $I$.

First, $I.\ {\bf e}^\phi$ is expanded by ${\bf e}^\perp$, ${\bf
e}^n$ and ${\bf e}^\psi$. But it should be orthogonal to ${\bf
e}^\psi$ from (\ref{rot1}): If ${\bf e}^\phi$ and ${\bf e}^\psi$
had components which mixes into each other by $I$, acting one more
$I$ on (\ref{rot1}) would still yield some component tangent to
$S^5$ in the right hand side, which is a contradiction. So we have
$I.\ {\bf e}^\phi=A\ {\bf e}^\perp+B\ {\bf e}^n$. Taking the norm
of this with ${\bf e}^\perp$, we obtain $A=-\cos\alpha$. This
proves the relation (\ref{rot3}), where the $+$ sign of $B$ is our
convention for the ${\bf e}^n$ direction. Then (\ref{rot2}) and
(\ref{rot4}) are  obtained from (\ref{rot1}) and (\ref{rot3}) by
applying the complex structure $I$.

\end{document}